----------
X-Sun-Data-Type: default
X-Sun-Data-Description: default
X-Sun-Data-Name: drele
X-Sun-Content-Lines: 1615

\documentstyle[12pt]{article}
\begin{document}
\begin{flushright}
FIAN-TD-8/95
\end{flushright}
\begin{center}
{\Large \bf Theoretical search for collective effects\\
in multiparticle production }\footnote{To appear in the volume of Uspekhi
Fizicheskikh Nauk (Physics-Uspekhi) dedicated to the centenary of Igor E.
Tamm (July issue, 1995).}
\end{center}

\medskip
\begin{center}
{\bf{I.M.~Dremin and A.V.~Leonidov}}\footnote {E-mail addresses:
dremin@td.lpi.ac.ru ; leonidov@td.lpi.ac.ru}
\end{center}
\begin{center}

{\it{Lebedev Physical Institute, Moscow, Russia}}

\end{center}
\medskip
\begin{center}
{\it{Abstract}}
\end{center}

           The properties of QCD vacuum and of the confinement of quarks
and gluons certainly influence the multiparticle production processes.
Some phenomenological attempts of the consideration of related collective
effects and the possibilities of their experimental detection are briefly
discussed in this review. We consider in particular the correlation
characteristics of pion systems, statistical and hydrodynamical analogies,
the problem of a phase transition from a quark-gluon plasma to a
 multipion state and the possible modifications of the evolution equations
of the quark-gluon jets. The presentation is somewhat simplified and could
be interesting for those only entering the field.

\begin{center}
{\bf{Contents}}
\end{center}

Foreword

1. Introduction\\
2. Early history of a problem. Statistical and hydrodynamical models\\
3. Correlation characteristics and methods of description of
multiparticle systems\\
4. Feynman-Wilson liquid. Statistical analogies\\
5. Quark-gluon plasma and multipion states\\
6. Collective effects in QCD jets\\
7. Discussion\\

References
\newpage

\begin{center}
{\bf{Foreword}}
\end{center}

       This review is not a traditional one, which is seen already from its
title. Usually a review contains a series of original papers providing
a relatively exhaustive presentation. Here we are attempting to describe
certain directions of theoretical search for the collective effects in
the particle interactions at high energies. Using the modern terminology
one could somewhat conventionally unify these directions and speak
about some phenomenological attempts to find the nonperturbative effects
which can not be described by the QCD perturbation theory. They are related
in particular to the QCD vacuum structure and the confinement of quarks and
 gluons. Although the applied methods and the level of the results vary
drastically, we have decided to give their brief description in the hope
that their comparison could be useful for the development of new ideas
and will lead to new suggestions. This is the reason for the presentation
not being detailed. We only mention the final results and for the detailed
explanations and calculations the reader should turn to the original papers.
 Our goal is to give a general perspective of several parallel approaches.

Additional motivation for this is a recollection of how lively Igor Evgenievech
Tamm, to whose centenary this volume is dedicated, always faced new ideas and
participated in their discussion. Moreover, the last paragraph of the
review is directly related to the work of I.E.~Tamm on the development of
electron-photon showers in the medium and on the Cherenkov radiation,
because there we discuss the hypothetical analogous effects in quark-gluon
jets.

\begin{center}
{\bf{1. Introduction}}
\end{center}

A collision of elementary particles or nuclei at very high energies is
usually accompanied by the production of many new particles. The multiparticle
production processes were first discovered in cosmic rays and then on the
accelerators. Their origin is a strong interaction of colliding particles
which, according to
the modern understanding (see [1-6]), is described as an interaction of
quarks and gluons in the framework of Quantum Chromodynamics (QCD).

Due to the asymptotic freedom property of QCD, when the coupling
constant becomes small at small distances,
an approach based on the QCD perturbation theory turned out to be extremely
fruitful in the study of hard processes with large momentum transfer.
Moreover, although it seems amazing, it turns out possible to describe
a number of properties of the soft processes by taking into consideration
 higher order contributions in perturbation theory and  conservation
laws.

At the same time the problem of translating from a language of quarks
and gluons into that describing the experimentally detectable hadrons
is still solved either at the axiomatic or at the model level. In the
first case one usually exploits a hypothesis of a local parton-hadron
duality, when the momentum distributions of partons and hadrons are
identified up to a multiplicative factor. This hypothesis is supported
both by theoretical considerations (a formation of colorless parton
clusters (preconfinement)) and by a number of experimental facts. The
model approach is usually used in Monte Carlo modelling of a transition
from partons to hadrons.

Despite of the impressive successes of the perturbative approach, one
should not forget that the problem of confinement of color objects,
quarks and gluons, still remains unsolved. In particular the influence
of confinement on the properties of hadrons in multiparticle production
processes could turn out to be nontrivial. One can not exclude, that
the collective aspects of the system's behavior as a whole and the
specific properties of quark-gluon QCD vacuum could show up.
For a theoretical description of such features the string models taking
into account confinehods of statistical physics (or, more
generally, macroscopic approaches) are usually more relevant than the
 perturbative calculations. The study of the collective excitation modes
of a hadron (quark-gluon) medium is a very complicated problem and is
actually at the infancy stage.

To give an example of the possible effects let us turn to the analogies
with the electrodynamics of a continuous media. An electric charge (e.g.,
 electron),  crossing a boundary of the two electrically neutral media,
having different refractivity indices, radiates photons.
 The radiation properties are determined precisely by this difference,
which describes the different collective properties of the two media.
At sufficiently high electron velocities the Cherenkov radiation might
also appear. Analogously, in the process of interaction of two hadrons
 or nuclei, the quarks that are hidden in one of them, can in principle
"see" the target as a whole, i.e. can radiate as in the case when
they pass through a color neutral medium. This is of course possible
only at relatively small momentum transfer.

At the same time one should not forget that this analogy is not complete.
The radiation of photon does not lead to a change of a charge of the
radiating particle. On the contrary, the color current changes in the
process of gluon emission, because the gluons are themselves color objects.
The space-time analysis of the process is especially relevant in that situation
[6].

Collective interactions can also be important in the case, when in the
process of interaction a quark-gluon plasma is formed. Such a possibility
is discussed already for a long time. The main criterion is
 an appearance of a  sufficiently high hadronic density in the
collisions of heavy nuclei.

The above-listed examples can only serve as guidelines in attempting
to apply the methods of statistical physics to the problem of multiparticle
production. Here it is a right place to stress that in the statistical
physics as well as in the study of multiparticle production processes
one often applies the same mathematical method. It is based on the calculation
of correlation characteristics in the phase space. The similarity of the
electro- and chromodynamics lagrangians allows to use certain analogies.
At the same time their difference leads to important distinctions, one
of them having been mentioned above. This shows up, in particular, in the
similarity and distinctions of the corresponding equations.

      We shall try to give a brief review of the theoretical attempts to
analyze the problem of collective effects in particle physics.
Unfortunately here it is difficult to choose some mainstream. The
development of these ideas is rather following several directions,
slowing down or completely stopping at certain obstacles. We shall
begin with a brief review of a history of a subject and then give
a more thorough discussion of the recent ideas.

\begin{center}
{\bf{2. Early history of a problem.\\ Statistical and hydrodynamical models}}
\end{center}

The processes in which a large number of particles is produced were
discovered in the cosmic ray showers more than 60 years ago. The first
attempt to describe them using the ideas of statistical physics and
hydrodynamics dates back to Heisenberg [7]. Somewhat later the analogous
 attempts were made by Wataghin [8]. An active discussion of this
approach began however after the appearance of the paper by Fermi [9]
where he introduced a particular statistical model of the multiparticle
production processes in nuclear collisions (a detailed description of the
model and its further development is given in the reviews [10,11]).
According to the main assumption of the model the process of
multiparticle production occurs via creation of a unique system, in
which there establishes a thermodynamical equilibrium. The distributions
of secondary particles are therefore described by the thermodynamical
formulae for blackbody radiation.  The legitimacy of such an assumption
was discussed many times (see ref. [12] and the reviews [10,11]). It
was agreed that if the models of this type are valid, this takes
place only in the domain of relatively low energies. At higher energies
(and higher multiplicities) the interaction of generated particles can
lead to such an expansion of this unified system that can be described
by the equations of hydrodynamics. This idea was put forward by Landau
[13] and was later widely exploited in the papers by many authors
(see the review [14]). Here it is worth mentioning, that the possibility
of applying the thermodynamical formulae for the description of an ensemble
of strongly interacting particles at realistic energies and multiplicities
is of course far from being evident. The critical analysis of the
basics of this approach can be found, for example, in the interesting
paper [15].

 According to the basic postulates of the statistical physics the Fermi
model used an assumption of a dominant role of a phase space in the
probability of a final state with $n$ particles, when a quantum-mechanical
matrix element is just a normalization factor. Let us write a general
expression for this probability:
\begin{equation}
P_n \sim \int |A_n|^2 \delta^4 (\Sigma p_i - \Sigma p_f) \prod_f d^3p_f,
\end{equation}
where $A_n$ are the transition amplitudes,  $p_i$  and  $p_f$ are
the four-momenta of initial and final particles. If we consider the
transition amplitudes $A_n$  as being independent of the final momenta
$p_f$, the integral over the phase space will factorially vanish
with growing $n$, because the mean particle momentum will be
proportional to $1/n$, i.e.
\begin{equation}
P_n \sim n^{-(3n-4)}
\end{equation}
at $n \rightarrow \infty$ and fixed total energy $E$ (the particle mass is
neglected). The factorial behavior of analogous type is known in
statistical physics as well.   However, if an assumption on the weak
amplitude dependence on the final momenta and particle number $n$
can be justified, this can happen only at low energies. With growing
energy and number of created particles the strong interaction in such
a quasiclassical system forces to describe its evolution rather as
the hydrodynamical expansion of a blob of the nuclear matter. In
the process of its expansion the temperature drops and the blob decays
into final particles. This idea immediately explains an important
experimentally known fact of the limited transverse momenta of
the particles. Let us mention, that the quantum field theory is still
unable to describe this phenomenon of the cut-off of large transverse
momenta.

If one tries to take this into account phenomenologically in the
general relation (1), it is necessary to note, that a "complete"
theory should provide this transverse momenta cut-off in the
integral (1) through the corresponding behavior of the amplitudes
$A_n$. If (according to experiment) this cut-off happens at finite
values of the transverse momenta, then instead of the estimate (2)
one gets a much slower decrease with growing $n$:
\begin{equation}
P_n \sim n^{-(n-2)}.
\end{equation}
In this case the phase space does not already  have a form of a $3n$-
dimensional sphere and takes a form of an $n$-dimensional cylinder in
this space
{\footnote{
In this connection one often speaks about a "cylindrical" phase space.
}}
(if one disregards the restrictions imposed by the conservation laws
that slightly deform this cylinder).

Another substantial factor, which is not accounted for in the statistical
approach to multiparticle production, is a rapid growth of a number of
field theory diagrams contributing to the amplitude $A_n$ at  large $n$.
For example, for the process of transition from two gluons to $n$ gluons
this growth even exceeds factorial one (see Table 1 in the review [16],
where the following numbers can be found: at $n=2$ one has 4 diagrams,
and at $n=8$ their number reaches already 10525900). Of course, the relative
phases of different contributions are really essential and it is currently
impossible to conclude how strongly the growth of a number of diagrams with
increasing $n$ affects the estimate (3). However, this factor can not be
neglected.

At high multiplicities this can lead, for example, to a change of an expansion
parameter in quantum chromodynamics, describing the strong interactions, from
the coupling constant $\alpha_s$ to its product at a factor of the order of
$n$,
and it will turn out that
\begin{equation}
|A_n|^2 \sim n! \alpha_s^n \sim(n \alpha_s)^n.
\end{equation}
As a result, the law of decreasing of $P_n$ at large $n$ will change from the
factorial (of Poisson type) distribution to the exponential one
 or to the one close to it. Precisely this
type of behavior is currently being discussed in quantum chromodynamics [17].

Here the appearance of a new expansion parameter is clearly seen when
one analyses the corresponding distributions over $n$ with the help of
their moments [18,19]. It is easy to see, that the processes with
high multiplicity $n$ determine the moments of multiplicity distribution
of a high order $q$. For example, only the processes having the
multiplicity $n>q$ contribute to the factorial moments of order $q$.
One finds that precisely the quantity $q \alpha_s^{1/2}$, corresponding
to the above-mentioned factor $n \alpha_s$, provides [18] an expansion
parameter for the solutions of equations on the generating functions
of $P_n$ distributions (see also the review [19]).

The appearance of this parameter actually supports the idea that at
high multiplicities the collective interaction effects become essential
{\footnote{
It is interesting that the resulting distributions are not infinitely
divisible (see [19]) which could also be related to the collective effects.
}}
although possibly not describable within the simplest statistical approach.
Nevertheless the mathematical methods applied in the studies of many-body
system (be it a statistical physics problem or a multiparticle production
process) is the same and is based on the analysis of the distributions
of particles and their correlations in the system under investigation.
Thus we begin with its description.

\begin{center}
{\bf{3. Correlation characteristics and methods of description\\
        of multiparticle systems}}
\end{center}

As it was already stressed, the approaches used in the analysis of the
properties of multiparticle systems are quite similar both in the
statistical physics (to mention one particular example, in laser physics)
and in the physics of multiparticle production at high energies. At fixed
given particle number $n$ in some phase space volume $\Omega$ the system
can be characterized by the probability density  $W_n (1, 2, ..., n)$,
where the arguments $1,2, \ldots ,n$ denote the corresponding (generally
speaking, multidimensional) coordinates of these particles in the phase
space. Such an approach is called exclusive.

 However, it is often
more convenient to use the so-called inclusive approach, where the total
number of particles in the system is not fixed and one considers just its
$q$ - particle characteristics. This is the way followed by the majority
of the experiments studying the multiparticle production. In this case
 the inclusive densities $\rho_q$ are related to the experimentally measured
inclusive differential cross-sections as follows:
\begin{equation}
\rho_q({\vec p_1}, \ldots ,{\vec p_q})={1 \over {\sigma_{in}}}
{d \sigma \over d^3p_1 \ldots d^3p_q},
\end{equation}
where $\sigma_{in}$ is a total cross-section of inelastic processes,
and a distribution over the $3$ - momenta of the final particles is
considered. Of course, the less detailed distributions are also used,
when one integrates over the certain momenta components. The integration
of the inclusive densities over the total phase space volume $\Omega$
gives the non-normalized factorial moments:
\begin{eqnarray}
{\tilde{F}}_q & \equiv & \int_{\Omega} d^3p_1 \ldots \int_{\Omega} d^3p_q
\rho_q({\vec p_1},\ldots  , {\vec p_q})
 =  \langle n(n-1) \ldots (n-q+1) \rangle \nonumber \\
 & = & \sum_0^{\infty} n(n-1) \ldots (n-q+1) P_n = \langle n \rangle^q F_q,
\end{eqnarray}
where $P_n$ is a probability to find an $n$-particle state of a
system (the so-called particle multiplicity distribution), and $F_q$
are the normalized factorial moments. The inclusive densities of
order $q$ are given by a sum of the exclusive density of the same
order and the integrals from the exclusive densities of higher order
over all variables that are not taken into account, i.e. in formal
notation
\begin{eqnarray}
 \rho_q(1, \ldots ,q) &=& W_q(1, \ldots ,q) \\ &+&
\sum_{m=1}^{\infty} {1 \over m!} \int_{\Omega}
 W_{q+m}(1,\ldots,q,q+1,\ldots,q+m)
\prod_{j=1}^m d (q+j). \nonumber
\end{eqnarray}
The inclusive densities $\rho_q$ are nonvanishing even if the particles
are statistically independent. Therefore (analogously to the cluster
decomposition in statistical mechanics) it is convenient to introduce
the so-called cumulant correlation functions $C_q$, which vanish  in
the case when the particles are completely statistically independent
[20-22]. The general formulae relating them to the inclusive densities
are quite cumbersome (see, e.g., [23-24]). Therefore we shall
reproduce only the formulae for the cases $q=2$ and $q=3$:
\begin{eqnarray}
C_2(1,2) & = & \rho_2(1,2)-\rho_1(1) \rho_2(2),  \\
C_3(1,2,3) & = & \rho_3(1,2,3)-\sum_{(3)} \rho(1) \rho_2(2,3)
+2 \rho(1) \rho(2) \rho(3),
\end{eqnarray}
where it is clearly seen that the contributions from lower order
correlations are subtracted from the higher order ones (the notation
$\sum _{(3)}$ stands for the summation over three possible particle
permutations).

All these results can be obtained in a unified form with the help of a
generating functional
\begin{equation}
G(z)=1+\sum_{q=1}^{\infty} {1 \over q!}
\rho_q(1,\ldots,n) z(1) \ldots z(q) \prod_{j=1}^q d(j),
\end{equation}
where $z(j)$ is a subsidiary function depending on ${\vec{p}}_j$. Then
\begin{equation}
\rho_q (1,\ldots,q) = {\delta^q G(z) \over
\delta z(1) \ldots \delta z(q)} | _{z=0},
\end{equation}
\begin{equation}
C_q (1,\ldots,q) = {\delta ^q \ln G(z) \over
\delta z(1) \ldots \delta z(q)} | _{z=0}.
\end{equation}
 At  $z $ = const   the generating functional becomes a generating
function of the multiplicity distribution, and the variational derivatives
 in (11), (12) become the ordinary ones, which lead to the non-normalized
factorial and cumulant moments of this distribution correspondingly.
The normalized factorial moments $F_q$ and the cumulants $K_q$  are
given by the formulae
\begin{equation}
F_q={1 \over {\langle n \rangle}^q} {d^q G(z) \over dz^Q} |_{z=0} ,
\end{equation}
\begin{equation}
K_q = {1 \over {\langle n \rangle} ^q}
{d^q \ln G(z) \over dz^q} | _{z=0} ,
\end{equation}
and the multiplicity distribution $P_n$ is
\begin{equation}
P_n = {1 \over n!} {d^n G \over dz^n} |_{z=-1} ,
\end{equation}
i.e. it is related to the generating function by the formula
\begin{equation}
G(z) = \sum_{n=0}^{\infty} (1+z)^n P_n .
\end{equation}
We see that differentiating the generating function one can compute
both inclusive and exclusive characteristics of a system depending
on a point $z$ in which the derivatives are taken.

As it is clear from all written above the direct computation of
Feynman diagrams within the perturbative approach in quantum field
theory is not well suited to the description of the multiparticle
production processes. On the one hand the number of diagrams grows
catastrophically with the increasing particle number, on another one
in the perturbative approach one considers the matrix elements of
the scattering operator for the transitions between the states having
a fixed number of particles. Therefore even if a calculation at
fixed multiplicity is made, one gets the exclusive quantities, whereas
according to (7) for computing inclusive characteristics one has to
perform the infinite summation of the exclusive probabilities. One
usually tries to bypass these difficulties [1,6] either by resumming
an infinite number of specially chosen (leading logarithms, etc.)
terms in the perturbative series, or using the equation for the
generating functions (their validity is again proved by comparing the
results with the same series calculated up to the definite order in coupling
constant). We should note that these approaches
turned out to be quite successful and fruitful in predicting and
describing many characteristics of the multiparticle production
processes in quantum chromodynamics (see [1,6]). However this
approach leaves unsolved the basic question whether the perturbation
theory can reproduce all the effects corresponding to an interaction
lagrangian in principle (in particular, the possible collective
effects). For shedding light on the latter it seems more promising
to exploit the analogies with statistical physics which we shall
try to discuss in the subsequent sections.

The inadequacy of the perturbative calculations of given subsets of
Feynman diagrams to the problem of finding the inclusive characteristics
of the multiparticle production processes is due to an overwhelming
complexity of a scattering operator in the basis of the eigenfunctions
of the particle number. Therefore the attempts are made to find a more
adequate representation. A well-known example of such kind can be the
laser radiation, where it is preferable to use the basis of the coherent
states. Below we shall consider the analogous attempts in the physics
of the multiparticle production processes. Of course, in this case
one often has to rely upon the specific model examples rather than
upon the initial QCD interaction lagrangian.

In the study of correlations in the multipion systems it also proved
fruitful to follow a direct analogy with hydrodynamics. The self-
similarity of vortices in hydrodynamics corresponds to a growth of
correlation functions at small scales. This led to the introduction
of the notions of intermittency and fractality, giving rise to the
powerlike growth of the factorial moments (13) with decreasing phase
space volume, in particle physics. We shall not discuss these questions
(a detailed review can be found in ref. [24]).

\begin{center}
{\bf{4. Feynman-Wilson liquid. Statistical analogies}}
\end{center}

Each individual event in the particle interactions at high energies can
be fully characterized by specifying (apart from the masses and quantum
numbers) the three-dimensional momenta of the secondary particles. The
endpoints of these vectors define a set of points, lying as a rule
in the above-mentioned cylindrical phase space. The correlations in the
position of these points are defined by the interaction lagrangian and
conservation laws. A large enough set of such events can be considered
as a statistical ensemble. In particular, Feynman [25] put forward
the analogy with a usual liquid by assuming the presence of   short-
range correlations in the ensemble. This idea was further developed by
Wilson [26], and after that the ensemble is called a Feynman-Wilson
liquid. Let us stress, that in contrast to the Fermi model here one
does not develop a new statistical interaction model, but attempts
to consider the statistical properties of the ensemble of particles
created in the interaction process at high energies using the analogy
with statistical mechanics. This topic is discussed in large number of
publications (from simple analogies to specific models, see refs.
 [11,27-37] and references therein). As the exclusive probabilities
$P_n$ characterize the volume of the phase space filled with the ensemble
of $n$ - particle events, they play the role of a partition function for
the canonical ensemble. The generation function defined by the relation
(16) (in a general case it is a functional) is analogous to a grand
canonical partition function. The role of the volume is played by the
maximal rapidity
\begin{equation}
Y = \ln {s \over m^2} ,
\end{equation}
 where $s$ is a total energy squared in the center of mass system (CMS), $m$ is
a particle mass.
It characterizes the size of the "cylinder" in the direction of the
longitudinal momentum. The variable $z$ is related to activity (or
a chemical potential $\mu=\ln (z+1))$ in statistical mechanics. Thus one
can calculate the "pressure" in such a liquid in a "thermodynamic
limit"  $Y \rightarrow \infty$ as
\begin{equation}
p(z) = \lim_{Y \rightarrow \infty} {\partial \ln G(z,Y) \over \partial Y}
\end{equation}
and its density as
\begin{equation}
\rho(z) = \lim_{Y \rightarrow \infty} {\partial \over \partial z}
{\ln G(z,Y) \over Y}
\end{equation}
and thus determine the equation of state (for details see [32-34]).
The analysis of experimental data on $e^+ e^-$ annihilation and
non-diffractive hadron-hadron processes at high energies within
this approach was carried out in ref. [34]. Using the formula (16)
and the data on $P_n$ one can calculate the generating function
 $G(z,Y)$ at different energies $Y$. A linear extrapolation to
 $Y \rightarrow \infty$ (see Fig. 1) allows to compute the slopes
for the different values of $z$ (i.e. for different chemical potentials).
The resulting dependence of pressure on the chemical potential (see
Figs. 2a and 2b) is especially interesting in relation to the
problem of the phase transition in such a system. The existence of a
phase transition should show up in the discontinuities of the pressure
derivatives over $z$. From Figs. 2a and 2b it follows, that in
$e^+ e^-$ processes such discontinuities are absent, whereas in the
hadron-hadron interaction the constancy of $p(z)$ at $z \leq 1$
turns into a rapid growth at large $z$. Such a difference in the
behavior of the pressure is interesting by itself, although a
conclusion on the existence of some phase transition would be
clearly premature.

 This problem could be approached from a different perspective [36,37]
by studying the behavior of the zeroes of a grand canonical partition
function in a complex $z$ plane as a function of a number of particles
in the system under consideration. The point is that at finite energy
one always has a certain maximally possible multiplicity $N$ . Therefore
the sum in the formula (16) terminates at this value of $N$, i.e.
 the generating functional becomes a polynomial from $z$ of degree $N$
and thus has $N$ zeroes in the complex plane. Lee and Yang have used
these properties of a partition function [36,37,21] and formulated a
method of locating the phase transition point by  finding a point
at the positive real $z$ axis to which the partition function zeroes
converge at large $N$. The analysis of the model $e^+ e^-$ events at the
energy $1000$ GeV performed in ref. [38]
has shown that the zeroes of the partition function lie on the circle
{\footnote{
More complicated configurations of zeroes were also discussed [39].
}}
in the complex $z$ plane and really converge to a real $z$ axis at
growing $N$. The real experimental data on $e^+ e^-$ and $p {\bar{p}}$
interactions at high energies also lead to the zeroes of a generating
function placed on the circle [40]. In the limiting point to which
the zeroes locations converge one should have a singularity of a
total generating function given by the formula (16) (i.e. at
 $N \rightarrow \infty$). The methods of investigating the character
of this singular point were proposed in ref. [41]. The question on
in which cases the appearance of such singularity corresponds to a
presence of a phase transition of some type in the particle production
processes still remains open.

There were many theoretical attempts to describe this transition at the
phenomenological level (see e.g. [29-33,35,42-47] and the paragraphs
4.2.4 and 4.2.5 in the review [24]). Like in all the phenomenological
 theories of critical phenomena the main problem is a choice of a
corresponding order parameter and such an expression for the
partition function, that at small values of this parameter the
analytically solvable Ginzburg-Landau hamiltonian is recovered
{\footnote{
Although the same form of the hamiltonian can be assumed at large
values of the order parameter too.
}}.
For the order parameter one usually takes either a certain mean field
(with the possible gaussian fluctuations) [29,33]  or a fluctuating
field of inclusive distributions [35]. In the latter case the order
parameter is local in the momentum space, which actually corresponds
to an intrinsically nonlocal approach in the usual space. In particular
even the "free" Ginzburg-Landau hamiltonian leads here to strong
correlations [35]. Within such a picture one can find the correlation
characteristics described in the paragraph 3 and also the details
of the behavior of pressure, density, etc. Although the indications
of the presence of singularities in these quantities and also in the
generating functionals were obtained, the specific values strongly
depend on the particular assumptions.

Thus we shall not present a detailed account of this series of papers,
and shall only briefly illustrate the idea at an example of the
coherent states [30], which by definition realize the eigenstates
of the annihilation operator
\begin{equation}
a({\vec{p}})|\Pi \rangle = \Pi ({\vec{p}}) | \Pi \rangle,
\end{equation}
where $\Pi ({\vec{p}})$ is some (generally speaking, complex) function
of ${\vec{p}}$. As has been already mentioned, the coherent states may
provide a more convenient representation for the operators, characterizing
the multiparticle production  process, than the particle number
representation (in analogy with laser physics). Of course, this does
not mean, that the created system is always in the coherent state.

In this case the inclusive density $\rho_q$ (5) takes the form
\begin{equation}
\rho_q(1, \ldots ,q) =
\int \delta \Pi |\Pi(1)|^2 \ldots |\Pi(q)|^2 e^{-F(\Pi)},
\end{equation}
where $F(\Pi)$ is an arbitrary functional analogous to the free
energy in statistical physics. Here the generating function, which
follows from (10) at $z$=const, is written as
\begin{eqnarray}
G(z) = 1+\sum_{q=1}^{\infty} {z^q \over q!} \int_{\Omega} \rho_q
\prod_{j=1}^q d(j) \nonumber \\
 = {1 \over N} \int \delta \Pi e^{-F(\Pi)}
e^{z \int d{\vec{p}} |\Pi ({\vec{p}})|^2},
\end{eqnarray}
where the normalization $N$ is fixed by the relation:
\begin{equation}
N = \int \delta \Pi e^{-F(\Pi)}.
\end{equation}
In the absence of the complete dynamical theory, which would allow to calculate
$F(\Pi )$, one is tempted to use a phenomenological ansatz. An analogy with the
phenomenological Ginzburg-Landau theory of
superconductivity leads to the expression for $F(\Pi)$ of the form
\begin{equation}
F(\Pi) = \int d{\vec{p}} [a|\Pi({\vec{p}})|^2
+ b |\Pi({\vec{p}})|^4 + c |{\partial \Pi \over \partial {\vec{p}}}|^2].
\end{equation}
The calculation of the thermodynamical quantities is then performed in
accordance with the formulae (18), (19). Different models correspond to
a different choice of the parameters $a,b,c$ and are considered in refs.
[29,30,33,35]. As usual, the phase transition point is the one  in which
the parameter $a$ goes through zero.

In this case it is possible to describe a large variety of the states
of a system, from coherent to chaotic ones. Therefore, the whole approach can
be considered either as a convenient parameterization of the data or as an
attempt to find out some dynamical features of processes. It is interesting to
note,
that the location of the singularity of the generating function is
sensitive to the parameter choice. It moves from infinity in the case
of a coherent state (Poisson distribution) approaching the point
$z=0$ (where the inclusive distributions are calculated, see eqs.
(13), (14)) with increasing chaoticity. At the same time the model
of coherent states can itself be modified taking into account the
squeezed and correlated states (see, e.g., the review [48]).

\begin{center}
{\bf{5. Quark-gluon plasma and multipion states}}
\end{center}

The problem of the phase transition became especially acute, when
there appeared an idea about the formation of the quark-gluon plasma
in those collisions, where a high energy density is reached. The natural
density scale in the hadronic matter is either the average nuclear or nucleon
density or that based on purely dimensional arguments, namely, on the
value of the QCD cut-off parameter. All these estimates give the values
of the same order lying in the range from 0.15 to 0.5 GeV/fm$^3$.
At much higher densities (exceeding 1-2 GeV/fm$^3$) one can expect
the appearance of a plasma of quarks and gluons. The estimates show, that
such energy densities can hopefully appear in nucleus-nucleus
collisions at high energies, or  in some fluctuations in hadron
interactions. As the QCD coupling constant decreases at high
temperatures and densities due to the asymptotic freedom property,
one can hope that in this domain the weakly interacting system of
quarks and gluons (plasma) can be theoretically described.

 The evolution
of quark-gluon matter with its subsequent transformation into hadrons
is one of the most cardinal problems of QCD and demands the application
of the methods of many-body theory. The high-temperature plasma behavior
in a certain frequency domain is described by perturbative QCD, but at
lower temperatures the description of the system is already given
using the nonperturbative calculations on the lattice in QCD (in
the vicinity of the phase transition) and the effective theories of meson
and baryon fields in the hadronic phase. It is necessary to stress,
that even in the high temperature domain the perturbative approach
is applicable only in the lowest orders in perturbation theory at
high frequencies and allows to calculate the energy density, pressure,
 plasmon damping, etc. [49,50]. However in the long-wavelength and
lower temperature domain the perturbation theory is not applicable.
In the phenomenological analysis, the influence
of the long-wavelength modes is often described with the help of a notion
of classical sources leading to the coherent states. At the same
time the short-wavelength excitations are considered to be responsible
for the quantum correlations, which in a thermalized system are
averaged with the Planck distribution (see, e.g., [51]). All the higher
correlations are often reduced to the products of the two-particle
ones (analogously to the coupled pair approximation [52]). The
relative contribution of these domains determines the relation
between the  chaotic and regular components of correlations, which
is also often used in the analysis of the experimental data (see, e.g.,
 [52]).

The lattice QCD calculations have revealed the presence of a phase
transition between the equilibrium  quark-gluon and hadron phases
at a temperature, which , according to the different estimates, is
of the order of 100 - 160 MeV [53-56]. Let us note, that such a temperature
was already mentioned as a maximally possible for the hadronic phase
in the statistical bootstrap models that were developed in the sixties
and seventies (see, e.g., [57]). The question concerning the order
of the transition is not completely settled. The conclusions often
depend even on the size of the lattice. However the latest investigations
[58-60] give some ground to think, that in the theory with two massless
dynamical quarks the transition is of the second order. At larger number of
massless quarks it seems to be of the first order. However when the mass
of the third (strange) quark grows, the transition is again becoming a second
order one already for the masses that are even somewhat smaller than
that of a strange quark. Thus the existence of such transition can
be taken as a starting point in the construction of the phenomenological
models.

The simplest situation from the point of view of its theoretical
description would be the formation of an equilibrium quark-gluon
plasma, then its expansion and accompanying cooling, the phase
transition to the equilibrium hadronic matter and its subsequent
expansion. This particular scenario lies in the foundation of the
majority of papers analyzing the possible detection of the
quark-gluon plasma in the experiments on heavy ion collisions.
The central question here is doubtlessly whether the quark-gluon
system, formed as a result of a heavy ion collision, has enough
time to thermalize. This determines whether we have a hydrodynamical
plasma expansion starting from some moment, or the process of
conversion into hadrons is essentially non-equilibrium one.

Unfortunately the unified approach describing all the stages of the
process still does not exist, and the description is quite fragmentary.
Most often one deals with the particular models of this or that stage.
Naturally this status of a problem will affect the subsequent presentation
in this review. We shall begin with the description of the hadronic
stage of a process. The dominant fraction of the created particles are
the pions, which in the first approximation can be treated as practically
massless and having relatively low energies in the center of mass system
of colliding high energy particles.

One of the simplest effects could be the influence of the laws of conservation
of quantum numbers on the properties of the pion system. Let us consider
for definitiveness the process of the collision of two high energy
nucleons, where in the final state, apart from the two nucleons,
 a certain number of pions has been created. The total isotopic spin of the
pion
system is limited, according to the conservation laws, to the values
$I=0,1,2$,  whereas in the general case it could take values up to
$n_{tot}$, where $n_{tot}$ is a total number of pions. This fact
significantly affects the pion charge distribution. Even if the
distribution in the total pion number $n_{tot}$ is poissonian (as it happens
when pions are produced by classical currents), it turns
out, that the separate distributions of the charged  ($n_{+}, n_{-}$) or
neutral
($n_0$) pions are much wider than the Poisson one (see Fig.3, which
is taken from ref. [61]).  The experimental confirmation of this fact
could be the "Centauro" events, where the charge particles are noticeably
dominating, or "anti-Centauro" ones with a large number of neutral
pions. Let us illustrate the idea by considering, following the
quasiclassical approach of ref. [62], the production of many pions in
a system with a zero isotopic spin. The characteristic initial assumption
is a possibility of describing the pion system that radiates the
final state pions as a classical field [61-66], i.e. that the number
of pions per phase space cell is assumed to be big. According to the
standard reduction formula, the amplitude of generation of $N$ pions
by the source $J$ equals
\begin{eqnarray}
A^{a_{1},\ldots,a_{n}} (k_1,\ldots,k_n) =
{\lim_{k^2_n \rightarrow m^2_{\pi}}} \int D\pi^a \int DJ^a W[J] \nonumber \\
\exp (iS[\pi]+i\int d^4x \pi^a J^a ) \prod_{n=1}^N \int d^4 x_n
e^{ik_n x_n} (-\partial^2_{x_n} - m^2_{\pi}) \pi^{a_n} (x_n),
\end{eqnarray}
where the functional integration over $J$ corresponds to the averaging
over the characteristics of the pion source. Let us notice, that the
radiation of a classical current exactly reproduces the language of
coherent states. The quasiclassical estimate of the amplitude, performed
in the assumption on the axial symmetry of the initial interaction and
on the isotopic symmetry of the pion system (i.e. of the zero total
isospin) [62], leads to the two characteristic conclusions that also
appear in other publications on this topic (in particular, such
conclusions were reached in the above-mentioned paper [61]).

Firstly, only the distribution over the total number of pions is
poissonian. The distributions over the number of neutral and
charged pions are much wider (see Fig. 3). For the state with zero isospin
a probability of finding $2n$ neutral pions in the system of $2N$
pions has a form [67,68]:
\begin{equation}
P(n,N) = {(N!)^2 2^{2n} (2n)! \over (n!)^2 2^{2N} (2N+1)!} .
\end{equation}
At large $n,N$ we have a characteristic distribution [69]:
\begin{equation}
P(n,N) \sim (n/N)^{-1/2} .
\end{equation}
Secondly, the conservation of the total isospin leads, for example,
to the specific angular correlations between the particles with different
charges. Let us give the characteristic formula for the correlation over
the azimuthal angle $\varphi$, obtained in [62] for the pions having
zero rapidity:
\begin{equation}
{(\sigma_{tot} {d\sigma^{\pi^+ \pi^-} \over dk_1 dk_2}
-{9 \over 10} {d\sigma ^{\pi^+} \over dk_1} {d\sigma ^{\pi^-} \over dk_2})
\over ({d\sigma ^{\pi^+} \over dk_1} {d\sigma^{\pi^-} \over dk_2})}
 ={3 \over 10} \cos ^2 (\varphi_1 - \varphi_2) .
 \end{equation}
The experimental verification of such predictions is to our opinion very
interesting.

 We have already seen that only taking into account the isospin conservation
within the framework of the quasiclassical approach can lead to the dramatic
changes of the naive ideas concerning the
character of particle multiplicity distributions. From the theoretical point of
view the most interesting calculations are the ones attempting to
use the effective pion lagrangians. In principle this gives a chance
of getting the model-independent predictions.

Before describing the corresponding results, we feel necessary to make
the following comment. The general feature of all the papers  using the
low-energy model lagrangians for the description of the distributions of
particles and their correlations in the multiparticle production processes
is an application of these lagrangians in the case, when the initial
energy of the collision is quite high. The question concerning the
possibility of neglecting the heavy modes and the interaction between
the modes is quite nontrivial. Due to the practically insurmountable
difficulties arising on the way of the complete solution of the problem,
it seems reasonable to analyze the consequences of the most radical
assumptions concerning the possible dynamical reasons causing the
sharp charge asymmetry of the events where a large number of pions
is generated.

In the recent literature a "disordered chiral condensate" was widely
discussed as a possible asymmetry source in the production of charged
and neutral pions in some fraction of the events ( in particular, in
the above-mentioned "Centauros") [70-77]. Let us remind, that the
transformation properties of mesons with respect to the chiral
transformations are determined according  to the corresponding properties
of the order parameter, which characterizes the spontaneous breaking
of chiral symmetry and given, for example, by the average from the
bilinear combination of quark fields
\begin{equation}
\Phi \sim \langle {\bar{q}}_L q_R \rangle,
\end{equation}
where $q_{R(L)}$ are the right (left) states of the massless quarks. For
investigating the character of the singularity of thermodynamical functions
in the vicinity of the phase transitions, it is desirable to find a solvable
model having the same symmetry. Then, according to the universality principle
based on the scale invariance near the critical point, the solutions of this
model will have the same set of singularities. In such an approach the order
parameter (29) can be rewritten in terms of a set of hadron fields having
the same symmetry. Therefore, the multipion states become related to the
massless
quark fields and quark-gluon plasma, i.e. there appears a hope to
describe a phase transition between these so differing  phases within such a
model. In the realistic case of two massless quark flavors the chiral field
can be written in the form
\begin{equation}
\Phi \sim \sigma \bullet {\bar{1}} + i {\bar{\tau}}
\bullet {\bar{\pi}},
\end{equation}
where $\sigma, {\bar{\pi}}$ are the real fields, ${\bar{\tau}}$ are the
standard Pauli matrices, the $\pi$ - meson fields ${\bar{\pi}}$ form
an isotriplet and $\sigma$ is an isosinglet.

In the case of $SU(3)$ - algebra the number of such fields is already $18$.
They form the scalar and pseudoscalar nonets.

 The dynamics of these degrees
of freedom can be described by the lagrangian of the linear $\sigma$ model
\begin{equation}
{\it{L}}   = {1 \over 2} (\partial \sigma)^2+
{1 \over 2} (\partial {\bar{\pi}})^2-V(\sigma,{\bar{\pi}}) ,
\end{equation}
where $V$ is a potential depending on the combination $\sigma^2+{\bar{\pi}}^2$.
In the standard version [78] the spontaneous symmetry breaking occurs via
the formation of the nonzero vacuum average of the field $\sigma$. The
isotriplet
fields remain massless, i.e. the pions are the goldstones of the chiral
group.

Let us now assume [71], that in some region of space the vacuum orientation
is different from the standard one, and, for example,
\begin{equation}
\langle \sigma \rangle = f_{\pi} \cos \theta, \;\;\;\;
\langle {\bar{\pi}} \rangle = f_{\pi} {\bar{n}} \sin \theta,
\end{equation}
 where $f_{\pi}=$ 93 MeV, and ${\bar{n}}$ is a unit orientation vector of
 ${\bar{\pi}}$.  Such an assumption presupposes a specific scenario of
the process, which is still not studied in details.

If the field $\Phi$ is isotropic with respect to a direction on the 3-
dimensional sphere in the 4-dimensional space with the angles defined as
\begin{equation}
(\sigma,\pi_3,\pi_2,\pi_1)=
(\cos \theta, \sin \theta \cos \phi, \sin \theta \sin \phi \sin \eta,
\sin \theta \sin \phi \cos \eta),
\end{equation}
then for the probability distribution of a given state $r \equiv \cos ^2 \phi$
we have:
\begin{equation}
\int_{r_1}^{r_2} dr P(r)=
{1 \over \pi^2} \int_{0}^{2 \pi} d \eta \int_0^{\pi}  d \theta \sin ^2 \theta
\int_{\arccos r_2^{1/2}}^{\arccos r_1^{1/2}} d \phi \sin \phi,
\end{equation}
as it was obtained for the first time in ref. [69] (see (3)):
\begin{equation}
P(r)={1 \over 2 {\sqrt{r}}},
\end{equation}
i.e. we have again returned to the formula (27). Thus the probability
of finding an event with a fraction of the neutral pions being smaller
than a certain given $r_0$ is
\begin{equation}
W(r<r_0) = {\sqrt{r_0}} ,
\end{equation}
which constitutes 10 \% even at $r_0$ = 0.01. The charge fluctuations in
such a system are much larger than those that would follow from the Poisson
 distribution, where the distributions are concentrated around $r$=1/3.

Let us notice that  the configuration (32) is not a solution of the
equations of motion and is therefore unstable. In order to give an
accurate quantitative estimate it is necessary to take into account the
presence of a domain wall separating the metastable configuration (32)
from the surrounding normal vacuum. Unfortunately such calculations have
not yet been performed.

To carry out more detailed investigations, it is necessary to specify
the source of the fluctuations of the chiral order parameter. In ref. [70]
this mechanism was considered in the context of a chiral phase transition,
when at increasing temperature the chiral symmetry restores at some
temperature $T_c$. Earlier it was established [55], that the QCD chiral
field corresponds to an $O(4)$-magnetic (which is natural for the choice
of the order parameter in the form of Eq.~(30)), for which one has a
second order phase transition. Therefore a most natural mechanism for
the generation of large scale fluctuations is a so-called "critical
slowing". In the vicinity of a critical point the relaxation time of the
long-wavelength degrees of freedom increases, and such fluctuations can
thus generate an isotopic disbalance, as the system's expansion can
"freeze" the  appearing isotopic inhomogeneities. It turns out, however,
that this mechanism is not effective. The reason is, that the tiny (on
the hadronic mass scale) current quark masses generate nevertheless a
sufficient pion mass $m_{\pi}$ (in fact, $m_{\pi} \sim T_c$), for the
chiral fluctuations to be indistinguishable from the thermal ones.

This has led the authors of [70] to considering the alternative possibility,
when the fluctuations that initially appear in the high-temperature
(unbroken) phase, subsequently evolve according to the zero temperature
equations of motion. Here the amplifying mechanism has quite remarkable
character [70]. In the spontaneously broken phase of a linear $\sigma$ model
with the condensate $v$ and the bare goldstone mass $\mu$, the zero mass
of the goldstone is provided by the relation
\begin{equation}
m^2_\pi = -\mu^2 + \lambda v^2  = 0 .
\end{equation}
Above the transition point one always has long-wavelength fluctuations,
for which $\langle \Phi^2 \rangle < v^2$, and during a certain time interval
the mass of these fluctuations is negative. Correspondingly the long-range
fluctuations will grow until the equations of motion restore the relation
$\langle \Phi^2 \rangle = v^2$. Such a situation could take place for a
rapidly expanding system. The resulting numerical solutions of
the corresponding evolution equations [70,72-75] are somewhat ambiguous.
We think that it would be premature to conclude, whether the fluctuations
of the chiral order parameter can be amplified to an observable scale.

In several papers [63,64,66,69] the problem was considered within a
nonlinear $\sigma$ model. Its lagrangian can be written in the form
(the discussion of the nonlinear $\sigma$-model can be found, for example,
in [79]):
\begin{equation}
{\it{L}} = {1 \over 2 f^2_{\pi}}
(\partial _\mu \Phi_a) (\partial^{\mu} \Phi_a)-
\lambda ({1 \over f^2_{\pi}} \Phi^2_a-1) ,
\end{equation}
where $\lambda$ is a Lagrange multiplier. In refs.~[63,69] the classical
solutions of the equations of motion of the nonlinear $\sigma$ model,
 describing the isotopic fluctuations of the chiral field, were found.
Unfortunately this has not led to more detailed predictions, although the
situation was clarified to some extent in [66,69].

An interesting attempt of obtaining the equations describing the quantum
fluctuations of the order parameter was recently made in refs.~[76,77].
The basic physical idea is that, as in the heavy nuclei collisions the
longitudinal expansion of hadronic matter in the pionization domain at
a scale of 8-10 fm dominates over the transverse one, the problem
becomes essentially two-dimensional. In ref.~[77] $SU(N_f)$ Wess-
Zumino-Novikov-Witten lagrangian for the chiral field
$U=\exp (2i {\bar{\tau}} {\bar{\pi}})$ was obtained as an effective 1+1 -
dimensional lagrangian.  In this model the fluctuations of the order
parameter can be computed exactly. In particular, for the two-point
rapidity correlations at a given proper time $\tau$ one gets
\begin{equation}
\langle U(\tau,Y) U^{\dagger} (\tau,Y' ) \rangle \sim
{1 \over ({\sqrt{2}} \tau)^{4 \Delta}} [\cosh (Y-Y')-1]^{2\Delta} ,
\end{equation}
where $\Delta={3/20}$. Such predictions could be tested
experimentally. Let us however note once again, that a method of
constructing the effective lagrangians which mixes the ideas originating
both form the high- and low- energy regime, does not allow to get
a consistent reduction. As a result the whole treatment is rather a
guess subject to experimental verification.

Where  should we look for the experimental signatures of the
existence of a chiral condensate and what are they? The hints of how
to answer the first part of the question follow from the fact, that
"Centauros" were found in cosmic rays but were not found in accelerator
experiments, although the Tevatron does already cover this energy domain.
If we do not ascribe the whole effect to the specific conditions of
 the registration of cosmic rays and their composition, we have to assume,
that the difference in the results originates from the fact, that in
 cosmic rays one studies the fragmentation region of large (pseudo)
rapidities, whereas in the accelerator experiments we were up to now studying
the pionization domain. According to the estimates of ref.~[71] the
"Centauro" type clusters should be looked for at Tevatron in the
pseudorapidity range in CMS of the order of 4 and higher. This is one
of the tasks of the T864 experiment. If the charge asymmetry is given
by the formula (27), it should be quite visible.

As we have mentioned, apart from a distinct signal of charge asymmetry
(27),(35) we also have got some conclusions on the correlation properties of
such
objects in azimuthal angle (28) and pseudorapidity (39). Moreover, the
electromagnetic decay modes of the hadronic resonances can be sensitive [73,74]
to the presence of such a condensate, because its orientation is
misaligned with that of the electroweak vacuum as defined by the Higgs
fields.

\begin{center}
{\bf{6. Collective effects in QCD jets}}
\end{center}

Let us turn now to QCD jets.From the point of view of field theory a typical
situation, in which
one uses the description in terms of the collective degrees of freedom,
arises when the system is considered as composed from a classical
subsystem (e.g., crystal, plasma, etc.) and corresponding quasiparticles
 (phonons, plasmons, etc.). In QCD it is natural to distinguish a domain
of small distances (quarks and gluons as high energy modes) and large
distances (hadronic modes). The high-energy phase is described by the
QCD perturbation theory, but in order to extract quantitative predictions,
 we should be able to estimate the contribution of the low-energy modes
to the characteristics that are traditionally computed within the
perturbation theory. One of the most interesting objects to look at are
here the hadron jets, generated by highly energetic quarks and gluons.

Let us begin with considering the possibility of the phenomenological
account of the presence of low-energy modes (confinement) on the evolution
 of the quark-gluon jets that give rise to hadron jets. In the simplest
approximation
(leading logarithms) the perturbative evolution of quark-gluon jets is
described by Gribov-Lipatov-Altarelli-Parisi equations (see, e.g., [1,6]).
The structure of these equations is very similar to the system of equations
describing the evolution of electromagnetic showers in matter [80].
The analogy becomes even more attractive, because the interaction with a
low-energy subsystem is inherently present in electromagnetic showers: these
are the inelastic losses on the ionization of the atoms of the medium by
the particles from the shower. We are pleased to note, that precisely this
problem was solved in the paper by S.Z.~Belenky and I.E.~Tamm [81],
whose centenary this issue of the journal is devoted to. We are
following the same line of investigations, but already in application
to quark-gluon jets.

Let us assume, that the influence of confinement (long-range modes
related to strings) on the evolution of hard quarks and gluons results
in the gradual loss of their energy going into the formation of pions.
This process resembles the loss of the electrons from the shower into
matter leading to the shower damping. As in the previous paragraph, we
get the amplification of the low-energy modes and the attenuation of the
hard component. In ref.~[82] a modification of the GLAP equations was
proposed, which treats the interaction with the low-energy modes in
analogy with the ionization losses. If we consider for simplicity only
a gluon subsystem, the modified equations take the form:
\begin{eqnarray}
{\partial D(E,{\tilde{Y}}) \over \partial {\tilde{Y}}}=
 2 \int_E^{\infty} {dE' \over E'}
 P^{GG}(E,E') D^G(E,{\tilde{Y}}) \nonumber \\
 - \int_0^E {dE' \over E}
 P^{GG}(E',E) D^G(E,{\tilde{Y}}) +
 \beta_G {\partial D(E,{\tilde{Y}}) \over \partial E},
 \end{eqnarray}
where $D^G(E,{\tilde{Y}})$ is a distribution function of gluons over the
energy $E$ and "depth"
\begin{equation}
{\tilde{Y}}={1 \over 2 \pi b}
\ln {\ln (Q^2/\Lambda^2) \over \ln (k^2/\Lambda^2)};\quad b={33 \over 12\pi},
\end{equation}
$Q^2,k^2$ are the initial and "current" invariant mass in a jet, $\Lambda$
is a cut-off parameter and $P^{GG}$ is a transition (gluon bremsstrahlung)
probability
\begin{equation}
P^{GG} = C_V x(1-x) [1+{1 \over x^2} +{1 \over (1-x)^2}]; \quad
x={E' \over E},
\end{equation}
$\beta_G$ is a dimensionful parameter determining the magnitude of the
nonperturbative inelastic losses. The solution of the Eq. (40) reveals the
maximum of the parton shower and rapid damping of the shower at low gluon
energies due to conversion of gluons into hadrons which influence is
phenomenologically described by the last term. No hadronization model was
attempted in [82,83]. However, the solution of the Eq.~(40) allows
 [83] to find the energy flow from a "hard" component to a soft one.
For the gluon jet with the initial energy $E_0$ the average energy of
 a hard component at the depth ${\tilde{Y}}$ is equal to
 \begin{eqnarray}
 \langle E({\tilde{Y}})  \rangle =
 E_0[1-{\beta_G \over E_0} {\sqrt{{{\tilde{Y}} \over 2c_V\ln (E_0/\beta_G)}}}
 I_1(2{\sqrt{2c_V {\tilde{Y}} \ln (E_0/\beta_G)}}) e^{-{\bar{a}} {\tilde{Y}}}],
 \end{eqnarray}
 where ${\bar{a}}=101/18$, $I_1$ is a modified Bessel function.

We see that the picture is qualitatively attractive, but for providing more
accurate predictions one should take into account the modification of the
initial equations due to kinematical and interference restrictions on the
perturbative evolution. It seems however quite probable, that the above
refinements will significantly alter the distribution function and
 correlations, but the formula for the energy losses (43) will most
probably remain essentially the same. The grounds for such a belief come
from the derivation of (43) in [83], where we essentially used the
energy conservation in the perturbative evolution. This phenomenological
approach to parton's hadronization is not complete since one should also
describe how many hadrons and with which characteristics appear as a result of
partons convolution. It is possible to do only with Monte Carlo models using
even more phenomenological parameters.

An attempt of a field-theoretical realization of the above-described
scheme was recently proposed in a series of papers [84,85]. The evolution of
parton showers both in space-time and in energy-momentum phase space is
analyzed. Some arguments in favor of rapid transition from partonic to hadronic
stage are given. Therefore, the idea of a unified
treatment of a high-energy subsystem and low-energy "thermostat" is
realized by  a literal addition of the corresponding lagrangians of
subsystem and interaction one:
\begin{equation}
{\it{L}}={\it{L}}_{QCD}+{\it{L}}_{eff}+{\it{L}}_{eff}^{int} ,
\end{equation}
where ${\it{L}}_{eff}$ is a low-energy lagrangian accounting for the
breaking of chiral and dilatational invariance [86,87] while the interaction
lagrangian describing the processes at the intermediate scales interpolates
between the two subsystems by phenomenologically introducing some effective
cut-offs imitating the rapid parton-hadron conversion within a very narrow
space-time region. In particular, if the gluon degrees of freedom are
considered only,
the lagrangian is chosen in the form:
\begin{eqnarray}
{\it{L}}=-{1 \over 4} {(G_{\mu \nu}^a})^2
+{1 \over 4} \xi (\chi) {(G_{\mu \nu}^a})^2  \nonumber \\
+{1 \over 2} (\partial_\mu \chi)^2
-b[{1 \over 4}  \chi_0^4+\chi^4 ln{\chi \over e^{1/4} \chi_0}],
\end{eqnarray}
where
\begin{equation}
\xi (\chi)=\theta(\chi) ({\chi \over \chi_0})^3 (4-{3\chi \over \chi_0}),
\end{equation}
 $\chi$ is a scalar glueball field, $G_{\mu \nu}^a$ is a gluon field strength,
and the interaction of the quantum gluon fields  $G_{\mu \nu}^a$ with the
"classical" $\chi$ has a form $\xi(\chi)G^2$. The non-vanishing gluon
condensate at large scales is denoted by $\chi _{0}$ and the parameter $b$ is
related to the conventional bag constant $B$ by $B=b\chi _{0}^{4}/4$. In the
chiral limit, the potential has a minimum when $\langle \chi \rangle =\chi
_{0}$ and equals the vacuum pressure $B$ at $\langle \chi \rangle =0$. From the
point of view of
the development of a gluon cascade the interaction lagrangian
 ${\it{L}}_{eff}^{int}$ leads to the appearance of the new vertices, i.e.
to a possibility of a cascade glueball formation. The natural requirement
to the modified lagrangian is a presence of an energy flow (from the quantum
component to the classical one) growing with increasing ${\tilde{Y}}$ that
reminds of the results of [83].
The numerical solutions of the corresponding kinetic evolution equations are
indeed demonstrating this property. The elaborated Monte Carlo program provides
characteristics of final hadrons in various reactions. The direction of
investigations in [84,85] is rather appealing even though there appear several
phenomenological parameters to be approved by later development. One can hope,
that already
in the nearest future the situation here will be clarified, and the problem
of the nonperturbative effects in the evolution of the quark-gluon
cascades  will reach the quantitative level.

Let us mention an interesting possible analogy with a collective
effect in the photon radiation. It is known, that the real part of an
elastic scattering amplitude of hadrons becomes positive at large
energies. In the terminology of classical physics this means, that the
 refractivity index of the hadronic medium exceeds one. In this case
the phase velocity of a colored charge in the hadron medium can be higher
than the speed of light. This can lead to the "color Cherenkov radiation"
 [88], which is analogous to the usual Cherenkov radiation, the theory
for which was developed by Tamm and Frank [89]. A characteristic feature
of such radiation will be its angular distribution with its typical "ring-like"
structure (see the review [90]
and the papers [91,92]). However, its intensity can be damped due to the
small size of the target hadron and the absorption in the hadronic medium
 (an imaginary part of the scattering amplitude). Besides a change of the
current in the course of emission of color objects could play its role (let
us remind that the photon is electrically neutral and,
unlike the gluon, does not change the emitting current at small recoils).

Nevertheless some events having a "ring structure" were observed in
cosmic rays [93], and for larger statistics the processing  of the
experimental data of the NA22 experiment gave indication [94] on  a
statistically significant contribution of such events. They show up
as peaks in the pion distributions at the rapidities  $|y| \sim 0.3$
in the center of mass frame. Several bands of energies are claimed [91,92] to
satisfy the requirements of "color Cherenkov radiation". The continuation of
the searches for the
possible manifestations of this effect would be desirable.

Let us now consider another aspect of the physics of hadron jets that is
also directly related to the considered collective effects. We are now
discussing the passage of a fast particle, which subsequently forms the
hadron jet, through the hadronic matter. A bright example of the effects
that could be expected in this connection is a sharp decrease of the
collisional energy losses in quark-gluon plasma near the phase transition
point (jet quenching [95]). Really, from the Bjorken estimate for the
energy losses of the parton in plasma at the unit length
\begin{equation}
{d {\it{E}} \over dx}  \sim 6 \alpha^2_s T^2 \ln {4ET \over M^2}
e^{-M/T} (1+{M \over T}),
\end{equation}
it follows that for the quite realistic values of $\alpha_s=0.2$,
$T=250$ MeV and $M=500$ MeV the energy losses of a parton at unit length
in plasma are 10 times smaller than the value of $1$ GeV/fm expected for
the usual hadronic matter, where the scale is determined by the string
tension. The accuracy of the answer directly depends on how well-studied
are the proper collective modes of a system. From classical plasma physics
it is well known that the interaction of external probe with the system
has a resonance character at the frequencies equal to the eigenfrequencies
of the system. In spite of the work on the study of the collisional losses
(see, e.g., [96]), the question concerning the quantitative estimate of the
effect is to our opinion open. In particular one does not know sufficiently
well
the spectrum of collective excitations of the quark-gluon plasma
 (recently, new modes were found in [97]).

Let us also discuss one more interesting effect illustrating the deep
diffferences between the effective nonabelian medium in different phases.
Considering the propagation of a color charge in the external random
chromoelectric field [98] a following formula for the change of the energy
of a fast test particle was obtained:
\begin{equation}
\langle {d{\it{E}} \over dt} \rangle =
{2\pi^{3/2} \alpha_s \langle {\bf{\cal{E}}}^2 \rangle
\tau_c \over m} {C_A \over N_c^2-1},
\end{equation}
where $\langle {\bf{\cal{E}}}^2 \rangle$ is a mean square of an external field
strength, $\tau_c$ is its characteristic correlation time, $m$  is a mass of
the
particle, $C_A$ is a corresponding Cazimir operator, $N_c$ is a number of
colors (a gaussian chromoelectric field correlator is assumed). Most
interesting is here even not a quantitative estimate of the possible
losses, but a fact that in the QCD vacuum one gets
 $\langle {\bf{\cal{E}}}^2 \rangle < 0$ (which is well known from the
consideration of heavy quarkonia [99]), whereas the fluctuations of the color
field in the QCD plasma naturally lead to
 $\langle {\bf{\cal{E}}}^2 \rangle >0$. Therefore we have a "stochastic
cooling" (deacceleration) in the vacuum in the hadronic phase and the
"stochastic heating" in e.g. QCD plasma, where in some way
the stochastic color fields were generated.

It is necessary to mention the recent growth of interest to the different
aspects of the physics of quark-gluon jets. We saw, how tightly the analysis
of these problems is related to the collective properties of the nonabelian
medium in which the jets propagate.

One of the most interesting is here a problem of coherent effects in the
induced gluon bremsstrahlung during the passage of a fast particle through
the nonabelian medium. We are speaking about the nonabelian analogue of
Landau-Pomeranchuk effect which is well known in electrodynamics [100-102].
The physical essence of this effect is a damping of a soft photon
radiation of a fast particle by a high energy particle due to a
coherent influence of many scatterers. It is interesting that the
theory of the electromagnetic effect was verified experimentally only
in 1993 [103]! For the physics of quark-gluon jets the question of the
intensity of  the induced bremsstrahlung radiation is a central one.
In particular, the above-mentioned estimates of the parton energy loss
 [96] are made under the assumption of the neglegiably small energy
losses at gluon radiation.  It may seem that a coherent radiation in the
QCD medium is simply impossible, because any radiated gluon changes the
color of a test parton, that should exclude any possibility of "using"
the multiple scattering for achieving the coherent action of a number of
scatterers on the bremmstrahlung gluon radiation. This point of view was
laid into a basis of calculations in [104], where for the energy losses
at unit length the authors obtained
\begin{equation}
-{d{\it{E}} \over dx} \sim const \cdot \alpha_s \mu^2,
\end{equation}
where $\mu$ is a screening length in the medium, which is cardinally
different from the electrodynamic result [100-102].

This statement was
recently objected in the paper  [105],
where the authors showed that a coherent action of a set of nonabelian
scatterers can be achieved by taking into account the regeneration
of the initial color by the gluon, emitted by the parent test parton.
Ideologically this situation resembles the color coherence effect in the
quark-gluon jets, where some gluons "see" only the total charge of a jet [6],
 which is analogous to the Chudakov effect [106], that is well known in
the physics of electromagnetic showers. An account of the above mechanism
the formula for the radiative energy losses in the nonabelian medium
 takes the form [105]
 \begin{equation}
 -{d{\it{E}} \over dx} \sim \alpha_s
 {\sqrt{{\it{E}} \mu^2 \over \lambda_g}}
 \ln ({{\it{E}} \over \lambda_g \mu^2}),
 \end{equation}
where $\lambda_g$ is a mean free path of gluons. The formula (50) differs
from the classical result of Migdal [101] by a logarithmic factors appearing
due to the correct treatment of collisions with a large momentum transfer.
 An analogous factor also appears in the corresponding electrodynamic
formula.

A problem of the parton energy losses in the nonabelian medium is of a
decisive importance for the description of the spectrum of jets in the
 collisions with nuclei.  The experimental study of hadron jets in such
collisions has already a relatively long history. In particular, there
exists a number of experimental results, which distinctly show the role
of a multiple scattering in the formation of two-jet configurations [107].
The continuation of the studying of this problem will doubtlessly
shed a light on the properties of the extremely complicated medium, in
which the jets are born and where they propagate.

Here it is tempting to point out an analogy with the cosmic rays, due to
 which we obtain information on the structure of fields in the intergalactic
space. It is possible, that a rich experience accumulated in the physics
of cosmic rays, will also help in the study of the "cosmic rays" of the
hadron physics, i.e. of the quark-gluon jets.

\begin{center}
{\bf{7. Discussion}}
\end{center}

We have tried to give a general description of the theoretical attempts
to find the collective effects in the multiparticle systems formed in
the collisions of high energy particles. In some way they are all having
a phenomenological character and are inspired by the hope to reveal certain
features of confinement and QCD vacuum structure and their role in the
multiparticle production process. The diversity of ideas, methods and
approaches underlines the absence of a unifying picture, but doubtlessly
is a necessary element on the way to its creation. We are sure that
 eventually it will appear. If we look only 60 years back, we shall see
 (see the article of E.L.~Feinberg in this issue), that the very notion
of a pion was still not accepted by everybody, and now we are trying
 to understand the properties of multipion states and their relation
to the quark and gluon degrees of freedom of an excited hadronic medium.
The absence of general conclusions means only that it is in the stage
of formation and dynamical development.

\begin{center}
{\bf{Acknowledgements}}
\end{center}

This work was supported by the Russian Fund for Fundamental  Research
under grant 93-02-3815 and, in part, by the NATO grant CRG930025.
\newpage

\begin{center}
{\bf{References}}
\end{center}

1. Andreev~I.V. (1981) Chromodynamics and hard processes at high energies
 (M.: Nauka) (in Russian).

2. Huang~K. (1982) Quarks, Leptons and Gauge Fields (Singapore: World
Scientific)

3. Ioffe~B.L., Lipatov~L.N., Khoze~V.A. (1983) Deep inelastic processes
(M.: Energoatomizdat) (in Russian)

4. Yndurain~F.J., (1983) Quantum Chromodynamics  (New York-Berlin-
Heidelberg, Tokyo: Springer Verlag)

5. Voloshin~M.B, Ter-Martirosyan~K.A. (1984)  Theory of gauge interactions
of elementary particles (M.: Energoatomizdat) (in Russian)

6. Dokshitzer~Yu.L., Khoze~V.A., Mueller~A.H., Troyan~S.I. (1991)
Basics of Perturbative QCD (Gif-sur-Yvette, France: Edition Frontiers)

7. Heisenberg~W., {\it{Zs.\ Phys.}}\ {\bf{101}}, 533 (1936);
   {\bf{113}}, 61 (1939)

8. Wataghin~G., {\it{Compt.\ Rend.}}\ {\bf{207}}, 358 (1938);
   {\it{Phys.\ Rev.}} {\bf{63}}, 137 (1943)

9. Fermi~E., {\it{Progr.\ Theor.\ Phys.}}\ {\bf{5}}, 570 (1950)
(see {\it{UFN}}\ {\bf{46}}, 71 (1952))

10. Belenky~S.Z., Maksimenko~V.M., Nikishov~A.I., Rosental~I.L.,
    {\it{UFN}}\ {\bf{62}}, 1 (1957)

11. Sisakyan~I.N., Feinberg~E.L., Chernavsky~D.S., {\it{Proc.\
 Lebedev\ Phys.\ Inst.}}\ {\bf{57}}, 164 (1972)

12. Pomeranchuk~I.Ya., {\it{DAN\ SSSR}}\ {\bf{78}}, 889 (1951)

13. Landau~L.D., {\it{Izv.\ AN\ SSSR,\ ser.\ fiz.}},\ {\bf{17}}, 51 (1953)

14. Feinberg~E.L., in "{\it{Advances in Theoretical Physics}}",
 {\it{Proc. of Landau birthday Symposium}}, ed. A.~L\"uther, Pergamon Press,
1988

15. Eggers~H.C., Heinz~U., {\it{Preprint}} {\bf{TPR-92-44}} (1992)

16. Mangano~M.L., Parke~S.J., {\it{Phys.\ Rep.}}\ {\bf{200}}, 301 (1991)

17. Dokshitzer~Yu.L., {\it{Phys.\ Lett.}}\ {bf{B305}}, 295 (1993)

18. Dremin~I.M., {\it{Phys.\ Lett.}}\ {\bf{B313}}, 209 (1993)

19. Dremin~I.M., {\it{Physics-Uspekhi}}\ {bf{37}}, 715 (1994)

20. Kahn~B., Uhlenbeck~G.E., {\it{Physica}}\ {\bf{5}}, 399 (1938)

21. Huang~K. (1963) Statistical Mechanics (John Wiley and Sons)

22. Mueller~A.H., {\it{Phys.\ Rev.}}\ {\bf{D4}}, 150 (1971)

23. Carruthers~P.A., {\it{Int.\ Journ.\ Mod.\ Phys.}}\ {\bf{A4}},
5587 (1989)

24. De~Wolf~E.A., Dremin~I.M., Kittel~W., {\it{UFN}}\ {\bf{163}},
3 (1993)

25. Feynman~R.P., {\it{Phys.\ Rev.\ Lett.}}\ {\bf{23}}, 1415 (1969)

26. Wilson~K., Preprint CLNS-131 (1970)

27. Hagedorn~R., {\it{Nucl.\ Phys.}}\ {\bf{B24}}, 93 (1970)

28. De~Groot~E.H., Ruijgrok~T.W., {\it{Nucl.\ Phys.}}\ {\bf{B27}},
45 (1971)

29. Scalapino~D.J., Sugar~R.L., {\it{Phys.\ Rev.}}\ {\bf{D8}},
2284 (1973)

30. Botke~J.C., Scalapino~D.J., Sugar~R.L., {\it{Phys.\ Rev.}}\
 {\bf{D9}}, 813 (1974); {\bf{D10}}, 1604 (1974)

31. Antoniou~N.G., Karanikas~A.I., Vlassopulos~S.D.F.,
 {\it{Phys.\ Rev.}}\ {\bf{D14}}, 3578 (1976);
 {\bf{D29}}, 1470 (1984)

32. Bjorken~J.D., {\it{Phys.\ Rev.}}\ {\bf{D27}}, 140 (1983)

33. Carruthers~P., Sarcevic~I., {\it{Phys.\ Lett.}}\ {\bf{B189}},
 442 (1987)

34. Hegyi~S., Krasnovszky~S., {\it{Phys.\ Lett.}}\ {\bf{B251}},
 197 (1990)

35. Dremin~I.M., Nazirov~M.T.
 {\it{Yad.\ Fiz.}} {\bf{55}}, 197, 2546 (1992)

36. Yang~C.N., Lee~T.D., {\it{Phys.\ Rev.}}\ {\bf{87}}, 404 (1952)

37. Lee~T.D., Yang~C.N., {\it{Phys.\ Rev.}}\ {\bf{87}}, 410 (1952)

38. De~Wolf~E.A., {\it{A note on multiplicity generating functions
in the complex plane}} (unpublished)

39. Andersson~B., in {\it{Proc. 24th Multiparticle Dynamics Symposium}},
 WSPC, Singapore, 1995

40.Gianini~G. (Private communication)

41.Dremin~I.M., {\it{Pisma\ v\ ZhETF}}\ {\bf{60}},
 757 (1994)

42. Satz~H., {\it{Nucl.\ Phys.}}\ {\bf{B326}}, 613 (1989)

43. Antoniou~N.G., Mistakidis~I.S., Diakonos~F.K.,
 {\it{Phys.\ Lett.}}\ {\bf{B293}}, 189 (1992)

44. Hwa~R.C., in {\it{Soft Physics and Fluctuations}}, WSPC,
 Singapore, 1994

45. Hwa~R.C., Pan~J., {\it{Phys.\ Rev.}}\ {\bf{C50}}, 2516 (1994)

46. Pan~J., {\it{Phys.\ Rev.}}\ {\bf{D46}}, R19 (1992)

47. The volume of {\it{Nucl.\ Phys.}}\ {\bf{A525}} (1991)

48. Dodonov~V.V., Dremin~I.M., Manko~O.V., Manko~V.I., Polynkin~P.G.,
 Preprint FIAN TD-03/95; {\it{Hadronic Journal}} (to be published)

49. Kapusta~J., {\it{Nucl.\ Phys.}}\ {\bf{B148}}, 461 (1979)

50. Pisarski~R.D., Braaten~E., {\it{Nucl.\ Phys.}}\ {\bf{B327}},
569 (1990); {\it{ibid}}\ {\bf{B339}}, 199 (1990)

51. Namiki~M., Muroya~S., in {\it{Proc. of the Int. Symp. on
High-Energy Nuclear Collisions and Quark-Gluon Plasma}},
p. 101, WSPC, Singapore (1991)

52. Carruthers~P., Sarcevic~I., {\it{Phys.\ Rev.\ Lett}}\ {\bf{63}},
1562 (1989)

53. Ukawa~A., {\it{Nucl.\ Phys.}}\ {\bf{A498}}, 227 (1989)

54. Engels~J., Fingberg~J., Karsch~F. et al.,
  {\it{Phys.\ Lett.}}\ {\bf{B252}}, 625 (1990)

55. Wilczek F., {\it{Int.\ Journ.\ Mod.\ Phys.}}\ {\bf{A7}},
 3911 (1992)

56. Polikarpov~M.I., {\it{UFN}}\ {\bf{165}}, N6 (1995)

57. Hagedorn~R., {\it{Suppl.\ Nuovo\ Cim.}}\ {\bf{3}},
 147 (1965); {\it{ibid}}\ {\bf{6}}, 311 (1965);
 {\it{Nuovo\ Cim.}}\ {\bf{A52}}, 1336 (1967)

58. Gottlieb~S., {\it{Nucl.\ Phys.}}\ {\bf{B20}}, 247 (1991)

59. Gausterer~H., Sanielevici~S., {\it{Phys.\ Lett.}}\ {\bf{B209}},
533 (1988)

60. Petersson~B., {\it{Nucl.\ Phys.}} [Proc. Suppl.]
{\bf{B30}}, 66 (1993);\\ F.~Karsch, E.~Laermann,
{\it{Rep.\ Progr.\ Phys.}}\ {\bf{56}}, 1347 (1993)

61. Andreev~I.V., {\it{Pisma\ v\ ZhETF}}\ {\bf{33}}, 384 (1981) (JETP Lett.
{\bf{33}}, 367 (1981))

62. Blaizot~J.P., Diakonov~D.I., {\it{On Semiclassical
Pion Production in Heavy Ion Collision}}, hep-ph/9307234

63. Andreev~I.V., {\it{Yad.\ Fiz.}}\ {\bf{17}}, 186 (1981)

63. Anselm~A.A., {\it{Phys.\ Lett.}}\ {\bf{B217}}, 169 (1989)

65. Bjorken~J.D., {\it{Acta\ Physica\ Polonica}}\ {\bf{B23}},
 561 (1992)

66. Blaizot~J.P., Krzywicki~A., {\it{Phys.\ Rev.}}\ {\bf{D46}},
 246 (1992)

67. Horn~D., Silver~R., {\it{Ann.\ Phys.}}\ (N.Y.)\ {\bf{66}},
 509 (1971)

68. Kowalski~K.L., Taylor~C.C., CWRUTH-92-6, hep-ph/9211282 (1992)

69. Anselm~A.A., Ryskin~M.G., {\it{Phys.\ Lett.}}\ {\bf{B266}},
 482 (1991)

70. Rajagopal~K., Wilczek~F., {\it{Nucl.\ Phys.}}\ {\bf{B379}},
 395 (1993); {\it{ibid}}\ {\bf{B404}}, 577 (1993)

71. Bjorken~J.D., Kowalski~K.L., Taylor~C.C.,
SLAC-PUB-6109 (1993)

72. Gavin~S., Gocksch~A., Pisarski~R.D., BNL-4965 (1993),
 hep-ph/9310228

73. Huang~Z., {\it{Phys.\ Rev.}}\ {\bf{D49}}, R16 (1994);\\
 Huang~Z., Wang~X-N., {\it{Phys.\ Rev.}}\ {\bf{D49}}, R4335 (1994)

74. Huang~Z., Suzuki~M., Wang~X-N., {\it{Phys.\ Rev.}}\ {\bf{D50}},
 R4335 (1994)

75. Gavin~S., M\"uller~B., {\it{Phys.\ Lett.}}\ {\bf{B329}}, 486 (1994)

76. Kogan~I.I., {\it{Phys.\ Rev.}}\ {\bf{D48}}, 3971 (1993)

77. Khlebnikov~S.Yu., {\it{Mod.\ Phys.\ Lett.}}\ {\bf{A8}},
 1901 (1993)

78. Gell-Mann~M., Levy~M., {\it{Nuovo\ Cimento}}\ {\bf{16}},
 705 (1960)

79. Lee~B.W. (1972) Chiral Dynamics (Gordon and Breach)

80. Belenky~S.Z. (1948) Showers in cosmic rays
 (M.: Gostechizdat) (in Russian)

81. Belenky~S.Z., Tamm~I.E., {\it{J.\ Phys.\ USSR}}\ {\bf{1}},
 177 (1939);
    Tamm~I.E. (1975) Scientific works, vol.2, p. 61
    (M.: Nauka) (in Russian)

82. Dremin~I.M., {\it{Pisma\ v\ ZhETF}}\ {\bf{31}}, 201 (1980)

83. Dremin~I.M., Leonidov~A.V., {\it{Yad.\ Fiz.}}\ {\bf{35}}, 430 (1981)

84. Geiger~K., {\it{Phys.\ Rev.}}\ {\bf{D46}}, 4965 (1992); {\bf{D49}}, 3234
(1994); {\bf{D51}}, 2345 (1995)

85. Ellis~J., Geiger~K., {\it{Preprints}} {\bf{CERN-TH.95-34;95-35}} (1995)

86. Lanik~J., {\it{Phys.\ Lett.}}\ {\bf{B144}}, 439 (1984)

87. Campbell~B.A., Ellis~J., Olive~K.A.,
{\it{Nucl.\ Phys.}}\ {\bf{B345}}, 57 (1990)

88. Dremin~I.M., {\it{Pisma\ v\ ZhETF}}\ {\bf{30}}, 152 (1979)

89. Tamm~I.E., Frank~I.M., {\it{DAN\ SSSR}}\ {\bf{14}}, 107 (1937);\\
    Tamm~I.E. (1975) Scientific works, vol.1, p. 68
    (M.: Nauka) (in Russian)

90. Dremin~I.M., {\it{El.\ Chast.\ At.\ Yad.}}\ {\bf{18}}, 79 (1987)

91. Ion~D.B., Stocker~W., {\it{Phys.\ Lett.}}\ {\bf{B273}}, 20 (1991);
    {\it{Ann.\ of\ Phys.\ (N.Y.)}}\ {\bf{213}}, 355 (1992); {\it{Phys.\
Lett.}}\      {\bf{B346}}, 172 (1995)

92. Hussein~M.T., {\it{Particle\ World}}\ {\bf{4}}, 15 (1994)

93. Apanasenko~A.V., Dobrotin~N.A., Dremin~I.M., Kotelnikov~K.A.,
 {\it{Pisma\ v\ ZhETF}}\ {\bf{30}}, 157 (1979)

94. Dremin~I.M., Zotkin~S.A., Lasaeva~P.E. et. al.,
 {\it{Yad.\ Fiz.}}\ {\bf{52}}, 840 (1990)

95. Gyulassy~M., Pl\"umer~M., {\it{Phys.\ Lett.}}\ {\bf{243}}, 432 (1990)

96. Thoma~M.H., Gyulassy~M., {\it{Nucl.\ Phys.}}\ {\bf{B351}}, 491 (1991);\\
    Braaten~E., Thoma~M., {\it{Phys.\ Rev.}}\ {\bf{D44}}, 1298 (1991);\\
    Mrowczynsky~S., {\it{Phys.\ Lett.}}\ {\bf{B269}}, 383 (1991)

97. Blaizot~J.-P., Iancu~E., {\it{Phys.\ Lett.}}\ {\bf{B326}}, 138 (1994)

98. Leonidov~A.V., {\it{Preprint}}\ {\bf{BI-TP\ 94/15}} (1994),
    to appear in {\it{Zeit.\ f.\ Phys.}}

99. Voloshin~M.B., {\it{Nucl.\ Phys.}}\ {\bf{B154}}, 365 (1979);\\
    Leitwyler~H., {\it{Phys.\ Lett.}}\ {\bf{B98}}, 447 (1981)

100. Landau~L.D., Pomeranchuk~I.Ya., {\it{DAN\ SSSR}}\ {\bf{92}}, 535 (1953)

101. Migdal~A.B., {\it{Phys.\ Rev.}}\ {\bf{103}}, 1811 (1956)

102. Feinberg E.L., Pomeranchuk~I.Ya., {\it{Nuovo\ Cimento\ Suppl.}}\
    {\bf{III}}, 652 (1956)

103. Perl~M.E., {\it{Preprint}}\ {\bf{SLAC-PUB-6514}} (1994)

104. Gyulassy~M., Wang~X.-N., {\it{Nucl.\ Phys.}}\ {\bf{B420}}, 583 (1994)

105. Baier~R., Dokshitzer~Yu.L., Peigne~S., Shiff~D.,
{\it{Preprint}}\ {\bf{LU TP 94-21; LPTHE 94/98; BI-TP 94/57}}

106. Chudakov~A.E., {\it{Izv.\ AN\ SSSR,\ Ser. fiz.}}\ {\bf{19}},
650 (1955)

107. Corcoran~M.D., Clark~R.K., Johns~K.A. et al.,
{\it{Phys.\ Lett.}}\ {\bf{B259}}, 209 (1991)

\newpage
\begin{center}
Figure Captions
\end{center}

 Fig. 1. The best approximation of the functions $\ln G(z,Y)/Y$ depending on
the energy $Y$ at various activities $z$ presented in ref. [34]. Let
us mention, that the data from the experiments in the $Z$ - boson region
are sufficiently well reproduced by the extrapolation of the straight lines
drawn in the figure (see [41]).

Fig. 2. The extrapolated according to the procedure of Fig. 1 to the point
 $Y^{-1}=0$  "experimental" values lead to the presented functional
dependencies of a pressure (calculated from the formula (18)) on the
activity $z$ for the cases a: electron-positron annihilation; b:
non-diffractive hadron-hadron collisions. The figure is taken from ref. [34].

Fig. 3. The distributions in the total number of pions and separately in
charged and neutral pions in the case of a pion system with zero isotopic
spin. The figure is borrowed from ref. [61]

\end{document}